\documentclass{elsart}
\usepackage{amsmath,bm}
\usepackage{graphicx}

\usepackage[normalem]{ulem}  
\usepackage[dvips]{color} 

\begin{document}

\begin{frontmatter}

\title{The neutron/proton ratio of squeezed-out nucleons and the high density behavior of the
nuclear symmetry energy}

\author[imp,grs,tamuc]{Gao-Chan Yong},
\author[tamuc]{Bao-An Li},
\author[tamuc,sjtu,nlhia]{Lie-Wen Chen}

\address[imp]{Institute of Modern Physics, Chinese Academy of Science, Lanzhou 730000, China}
\address[grs]{Graduate School,Chinese Academy of Science, Beijing 100039, P.R. China}
\address[tamuc]{Department of Physics, Texas A\&M University-Commerce, Commerce, TX 75429, USA }
\address[sjtu]{Institute of Theoretical Physics, Shanghai Jiao Tong University, Shanghai 200240, China}
\address[nlhia]{Center of Theoretical Nuclear Physics, National Laboratory of Heavy Ion Accelerator, Lanzhou 730000, China }

\begin{abstract}
Within a transport model it is shown that the neutron/proton ratio
of squeezed-out nucleons perpendicular to the reaction plane,
especially at high transverse momenta, in heavy-ion reactions
induced by high energy neutron-rich nuclei can be a useful tool
for studying the high density behavior of the nuclear symmetry
energy.
\end{abstract}


\end{frontmatter}

The equation of state (EOS) of neutron-rich nuclear matter is not
only important for nuclear physics but also crucial for many
phenomena and processes in astrophysics and cosmology, see, e.g.,
Refs. \cite{lat01c,steiner05a} for a recent review. Heavy-ion
reactions induced by neutron-rich nuclei, especially radioactive
beams, provide a unique opportunity to constrain the symmetry energy
term in the EOS of isospin asymmetric nuclear matter
\cite{ireview,ibook,baran05}. Though considerable progress has been
made recently in determining the density dependence of the nuclear
symmetry energy at sub-normal densities, probing the high density
behavior of the nuclear symmetry energy remains a major challenge.
As an illustration of the current situation, in Fig.\ 1 several
typical theoretical model predictions \cite{diep,fuchs06,zuo02} are
compared with some phenomenological constraints obtained from model
analyses of experimental data (those labeled $x=0$, $x=-1$ and
FSU-Gold). The constraints labeled $x=0$ and $x=-1$ were extracted
recently from studying isospin diffusion in the reaction of
$^{124}$Sn +$^{112}$Sn at $E_{beam}/A=50$ MeV within a transport
model \cite{ls03,mbt,chen04,li05}. For this particular reaction the
maximum density reached is about $1.2\rho_0$. Moreover, it was shown
that the neutron-skin thickness in $^{208}$Pb calculated within the
Hartree-Fock approach using the same underlying Skyrme interactions
as the ones labeled $x=0$ and $x=-1$ is consistent with the
available experimental data \cite{steiner05b,ba0511,chen05nskin}.
The symmetry energy labeled as FSU-Gold was calculated within a
Relativistic Mean Field Model (RMF) using a parameter set such that
it reproduces both the giant monopole resonance in $^{90}$Zr and
$^{208}$Pb, and the isovector giant dipole resonance of $^{208}$Pb
\cite{piek05}. It is also interesting to mention that the constraint
obtained recently from isoscaling analyses is consistent with the
FSU-Gold and the $x=0$ case \cite{shetty}. These results all
together represent the best phenomenological constraints available
on the symmetry energy at sub-normal densities. The various
predictions at supra-normal densities diverge widely although they
are all close to the existing constraints at low densities. The
lines labeled $x=0$, $x=-1$ and FSU-Gold are all extended into the
supra-saturation regions where they are not constrained by any
experimental data at all. The situation summarized in Fig.\ 1
signifies clearly the need for more work to determine the high
density behavior of the nuclear symmetry energy.
\begin{figure}[th]
\begin{center}
\includegraphics[width=0.85\textwidth]{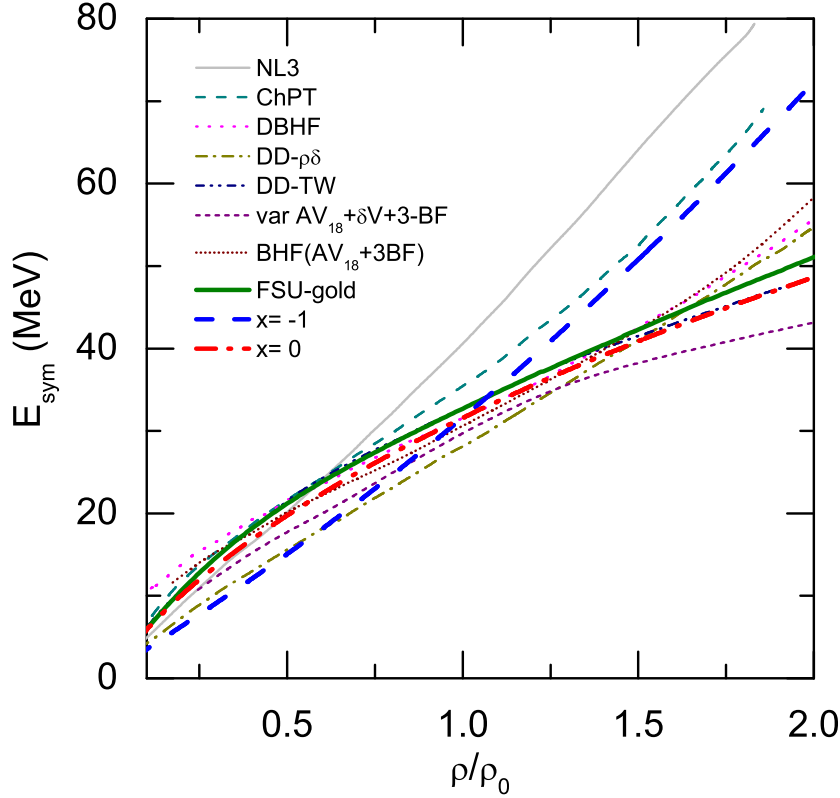}
\end{center}
\caption{(Color online) Density dependence of the nuclear symmetry
energy using the MDI interaction with $x=0$ and $x=-1$ and other
many-body theories predictions (data are taken from
\protect\cite{fuchs06,zuo02,piek05}).} \label{sym}
\end{figure}

Of particular interest is to identify experimental observables
that are sensitive to the high density behavior of the nuclear
symmetry energy. However, it is very challenging to find such
observables since the high density phase is formed only
transiently in heavy-ion reactions. Moreover, most hadronic
observables are affected by both the isospin symmetric and
asymmetric parts of the EOS at all densities throughout the whole
dynamical evolution of the reaction. Thus rather delicate
observables have to be selected to probe the high density behavior
of the nuclear symmetry energy. The $\pi ^{-}/\pi^{+}$ ratio
\cite{ba02a,gai04,qli05a,qli05b}, the neutron-proton differential
flow \cite{li00,yong062} and the $K^{0}/K^{+}$ ratio
\cite{ditoro06} are among the most promising observables
identified so far for this purpose. In this Letter, it is shown that the
neutron/proton ratio of squeezed-out nucleons, especially at high
transverse momenta, in heavy-ion reactions induced by high energy
neutron-rich nuclei is another useful, complementary but more
direct tool for studying the high density behavior of the nuclear
symmetry energy.

Our study is based on the transport model IBUU04. Details of the
model and its applications in studying the density dependence of
the nuclear symmetry energy can be found in Refs.
\cite{ba97a,ba04a}. The most important input relevant to this work
is the momentum- and isospin-dependent single nucleon potential (MDI) \cite%
{das03}, i.e.
\begin{eqnarray}\label{mdi}
U(\rho ,\delta ,\mathbf{p},\tau ) &=&A_{u}(x)\frac{\rho _{\tau ^{\prime }}}{%
\rho _{0}}+A_{l}(x)\frac{\rho _{\tau }}{\rho _{0}}  \nonumber \\
&&+B(\frac{\rho }{\rho _{0}})^{\sigma }(1-x\delta ^{2})-8x\tau \frac{B}{%
\sigma +1}\frac{\rho ^{\sigma -1}}{\rho _{0}^{\sigma }}\delta \rho
_{\tau
^{\prime }}  \nonumber \\
&&+\frac{2C_{\tau ,\tau }}{\rho _{0}}\int d^{3}\mathbf{p}^{\prime }\frac{%
f_{\tau }(\mathbf{r},\mathbf{p}^{\prime
})}{1+(\mathbf{p}-\mathbf{p}^{\prime
})^{2}/\Lambda ^{2}}  \nonumber \\
&&+\frac{2C_{\tau ,\tau ^{\prime }}}{\rho _{0}}\int d^{3}\mathbf{p}^{\prime }%
\frac{f_{\tau ^{\prime }}(\mathbf{r},\mathbf{p}^{\prime })}{1+(\mathbf{p}-%
\mathbf{p}^{\prime })^{2}/\Lambda ^{2}}.  \label{potential}
\end{eqnarray}%
The detailed values of the parameters can be found in Ref. \cite%
{chen04,li05,das03}. The parameter $x$ in Eq. (\ref{mdi}) was
introduced to mimic various forms of the density dependence of the
symmetry energy. For a given value $x$, one can readily calculate
the symmetry energy $E_{\text{sym}}(\rho )$ as a function of density
\cite{das03}. As we have stressed earlier, while the symmetry energy
at sub-normal densities is constrained approximately between $x=0$
and $x=-1$, there is still no experimental indication about the
symmetry energy at supra-normal densities. Assuming the continuity
of the model to higher densities we use in the present work the MDI
interaction with $x=0$ and $x=-1$ at all densities.

It is well known that the squeeze-out of nuclear matter in the
participant region perpendicular to the reaction plane occurs in
noncentral heavy-ion collisions. In mid-central collisions, high
density nuclear matter in the participant region has larger
density gradient in the direction perpendicular to the reaction
plane. Moreover, in this direction nucleons emitted from the high
density participant region have a better chance to escape without
being hindered by the spectators. These nucleons thus carry more
direct information about the high density phase of the reaction.
They have been widely used in probing the EOS of dense matter,
see, e.g., Refs.
\cite{greiner,bert88,cas90,aich91,Reisdorf97,danie02} for a
review. We explore here whether the squeeze-out of nucleons can be
used to constrain the high density behavior of the nuclear
symmetry energy. As an example, we consider here the reaction of
$^{132}$Sn+$^{124}$Sn at a beam energy of $400$ MeV/nucleon and an
impact parameter of 5 fm. In this reaction the maximal baryon
density reached is about twice the normal nuclear matter
density\cite{yong062}.
\begin{figure}[th]
\begin{center}
\includegraphics[width=0.85\textwidth]{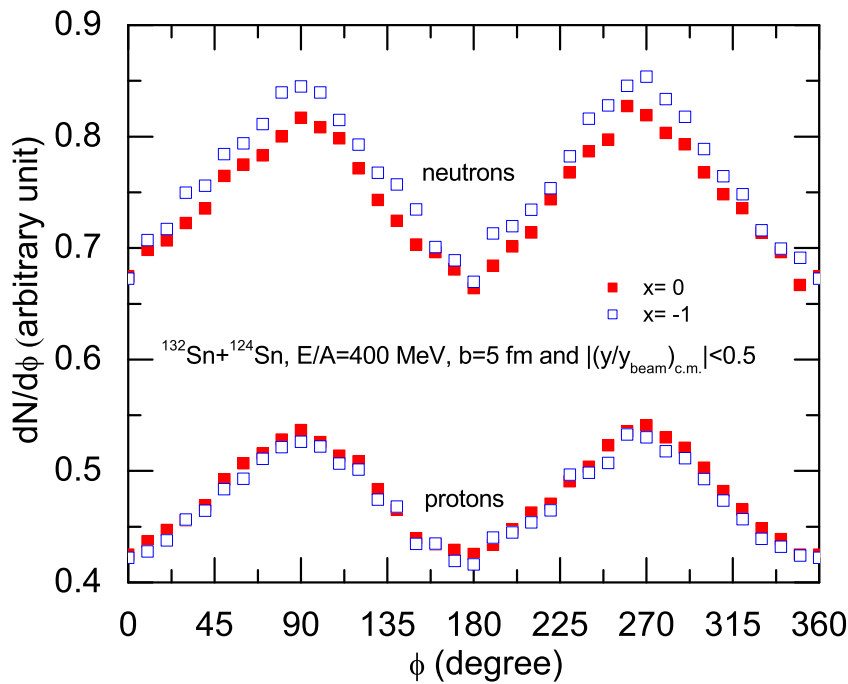}
\end{center}
\caption{(Color online) Azimuthal distribution of midrapidity
nucleons emitted in the reaction of $^{132}$Sn+$^{124}$Sn at an
incident beam energy of $400$ MeV/nucleon and an impact parameter
of $b=5$ fm.} \label{degree}
\end{figure}

Shown in Fig.\ \ref{degree} are the azimuthal distributions of free
nucleons in the midrapidity region ($|(y/y_{beam})_{c.m.}|<0.5$). A
preferential emission of nucleons perpendicular to the reaction
plane is observed clearly for both neutrons and protons as one
expects. Most interestingly, neutrons emitted perpendicular to the
reaction plane show clearly an appreciable sensitivity to the
variation of the symmetry energy compared to protons. This is mainly
because the symmetry potential is normally repulsive for neutrons
and attractive for protons. For the latter, the additional repulsive
Coulomb potential works against the attractive symmetry potential.
Overall, one thus expects the neutron emission to be more sensitive
to the variation of the symmetry energy. Since the symmetry
potential is relatively small compared to the isoscalar potential,
it is always necessary and challenging to find obervables that are
delicate enough to be useful for extracting information about the
symmetry potential/energy. Fortunately, the squeeze-out of neutrons
appears to be promising.
\begin{figure}[th]
\begin{center}
\includegraphics[width=0.85\textwidth]{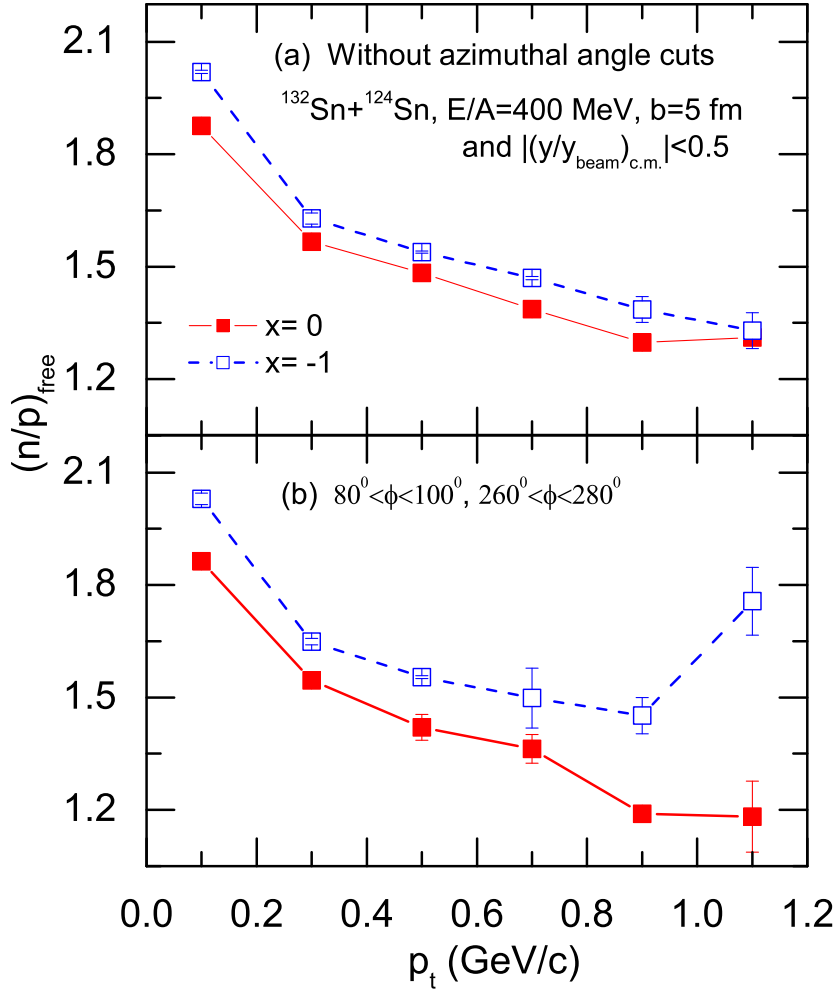}
\end{center}
\caption{(Color online) Transverse momentum distribution of the
ratio of midrapidity neutrons to protons emitted in the reaction
of $^{132}$Sn+$^{124}$Sn at the incident beam energy of $400$
MeV/nucleon and impact parameter of $b=5$ fm. For the lower panel,
we make an azimuthal angle cut of $%
80^{\circ}<\protect\phi <100^{\circ}$ and $260^{\circ}<\protect\phi
<280^{\circ}$ to make sure that the free nucleons are from the
direction perpendicular to the reaction plane, i.e., to analyze the
squeezed-out nucleons only.} \label{rnp}
\end{figure}

Since our main focus in this work is to probe the high density
behavior of the symmetry energy, we would like to avoid as much as
possible all remaining uncertainties associated with the EOS of
symmetric nuclear matter. For this purpose we examine in the lower
window of Fig.\ \ref{rnp} the transverse momentum dependence of the
neutron/proton (n/p) ratio of midrapidity nucleons emitted in the
direction perpendicular to the reaction plane. It is known from
previous studies \cite{ireview,ba97a} that the n/p ratio is
determined mostly by the density dependence of the symmetry energy
and almost not affected by the EOS of symmetric nuclear matter. It
is interesting to see in the lower window that the symmetry energy
effect on the n/p ratio increases with the increasing transverse
momentum $p_t$. At a transverse momentum of 1 GeV/c, the effect can
be as high as 40\%. The high $p_t$ particles most likely come from
the high density region in the early stage during heavy-ion
collisions and they are just more sensitive to the high density
behavior of the symmetry energy. Without the cut on the azimuthal
angle, the n/p ratio of free nucleons in the midrapidity region is
shown in the upper window. This ratio is much less sensitive to the
symmetry energy in the whole range of transverse momentum. It is
worth mentioning that the n/p ratio of free nucleons perpendicular
to the beam direction in the CMS frame in $^{124}$Sn+$^{124}$Sn
reactions at 50 MeV/nucleon was recently measured at the NSCL/MSU
\cite{msu06}. This measurement was useful for studying the density
dependence of the symmetry energy at sub-normal densities.

Compared to other potentially powerful probes of the symmetry energy
at supra-normal densities, such as, the $\pi ^{-}/\pi ^{+}$ and
$K^{0}/K^{+}$ ratios, the n/p ratio of squeezed-out nucleons carries
directly information of the symmetry potential/energy since it acts
directly on nucleons. Pions and kaons are mostly produced through
nucleon-nucleon and pion-nucleon inelastic scatterings, they thus
carry indirectly and often secondary or even higher order effects of
the symmetry energy \cite{li05b}. Moreover, nucleonic observables
such as the n/p ratio are essentially free of uncertainties
associated with the production mechanisms of pions and kaons.

While it is very tough to measure neutrons, both the transverse flow
and the squeeze-out of neutrons were measured at the BEVALAC by
Madey et al.\cite{bevalac1,bevalac2} and at the SIS/GSI by the TAPS
and the Land collaborations\cite{gsi1,gsi2,gsi3}. The measurements
were accurate enough to extract reliable information about the EOS
of symmetric nuclear matter and the reaction dynamics. The analyses
of the experimental data and the associated theoretical
calculations, see, e.g., Refs. \cite{bass,lar00}, however, have all
focused on extracting only information about the EOS of symmetric
nuclear matter without paying any attention to the symmetry energy.
In all of these experiments, it was essential to measure
simultaneously charged particles together with neutrons. To study
the symmetry energy at high densities using the n/p ratio of
squzzed-out nucleons, similar experimental setups are necessary.
Especially, to construct the reaction plane, $4\pi$ charged particle
detectors are necessary. For determining the momenta of neutrons,
the TOF (Time of Flight) of neutrons can be measured with neutron
walls or other neutron detectors. The squeeze-out of nucleons can
then be studied with respect to the reaction plane determined by
using the charged particles on the event-by-event basis.

The symmetry energy effects on the n/p ratio of squeezed-out
nucleons are large enough to be measured with some of the existing
detectors\cite{gary}. This optimistic view and the past success in
studying neutron squeeze-out make us feel confident that the
predicted effects can be studied realistically. Especially, with the
new development in detector technologies, the next generation of
detectors for both charged particles and neutrons are being planned
and/or constructed. For instance, the Modular Neutron Array (MoNA)
or the neutron walls existing at the NSCL/MSU may be coupled with
one of the available charged particle detectors there. However, it
is beyond the scope of this work to estimate the technical
requirements for the detectors to measure the predicted symmetry
energy effects on the n/p ratio of squeezed-out nucleons.
Nevertheless, it is still exciting to mention that the nuclear
reaction community is currently considering the construction of a
new TPC (Time-Projection-Chamber) for studying reactions induced by
high energy radioactive beams\cite{lynch}. To study the n/p ratio of
squeezed-out nucleons in these reactions, an advanced neutron
detector must be used together with the TPC. The present study adds
to the importance of constructing the TPC and an advanced neutron
detector at the site of an isotope science facility capable of
providing high energy radioactive beams.

In summary, it has been a challenging task for the intermediate
energy heavy-ion community to identify experimental observables that
are sensitive to the high density behavior of the nuclear symmetry
energy. Within a transport model, we found that the neutron/proton
ratio of squeezed-out nucleons perpendicular to the reaction plane,
especially at high transverse momenta, is such an observable.
Compared to other potential probes identified earlier in the
literature, the n/p ratio of squeezed-out nucleons is complementary
but carries more direct information about the symmetry energy at
high densities. The sensitivity to the high density behavior of the
nuclear symmetry energy observed in the n/p ratio of squeeze-out
nucleons is probably the highest found so far among all observables
studied within the same transport model.

We would like to thank Dr. Wei-Zhou Jiang, Dr. Plamen Krastev, Dr.
Gary Westfall and Dr. S.J. Yennello for helpful discussions. The
work was supported in part by the US National Science Foundation
under Grant No. PHY-0652548, the Research Corporation, the National
Natural Science Foundation of China under Grant Nos. 10575071 and
10675082, MOE of China under project NCET-05-0392, Shanghai
Rising-Star Program under Grant No. 06QA14024, and the SRF for ROCS,
SEM of China.

\end{document}